\documentstyle[12pt]{article}
\newtheorem{pr}{Proposition}

\begin{document}

\begin{center}
{\Large \bf On the braided Fock spaces}
\end{center}

\begin{center}
A.V.Mishchenko

Bogolyubov Institute for Theoretical Physics,
National Academy of Sciences of Ukraine, \\
252143 Kiev, Ukraine \\
e-mail: \\
olmishch@olinet.isf.kiev.ua \\
or \\
olmishch@ap3.gluk.apc.org
\end{center}

\begin{abstract}
Framework for constructing Fock spaces associated either
with certain
solutions of the quantum Yang-Baxter equation
or with infinite dimensional Hecke algebra is presented.
For the former case, the quantum deformed oscillator algebra
associated with the solution of the quantum Yang-Baxter
equation is found.
\end{abstract}

During last decade much attention have been paid to the quantum
(q-deformed) groups and algebras (both will be called
quantum groups below)
as well as to their
applications in such diverse branches
of theoretical and mathematical physics
as conformal field theory, integrable models, nuclear physics,
statistical mechanics, knot theory and topological field
theory (see, e.g., \cite{chang} and references therein).
For the first time they appeared in fact as the
main ingredient of quantum inverse scattering
method \cite{fad1,fad2} and
then were interpreted as Hopf algebras in
\cite{drinf1,drinf2,jimbo} (see also \cite{woron} for
the quantum matrix group interpretation). There exist various
models describing systems of
either one particle or several distinct
particles, which possess quantum symmetry, i.e., are symmetrical
with respect to the action of quantum group
(see, e.g., \cite{chang}, \cite{wess1} - \cite{kibler}).
Unfortunately, the problem of description of a
quantum symmetric (identical)
multiparticle system is not solved completely yet.
It is well known that quantum groups are deeply
connected with the braid group, which
substitutes usual symmetric group in this
case \cite{madj,fad3,katriel} (see also \cite{resh1}).
Therefore, one can expect that
to construct a multiparticle system possessing
quantum symmetry, the braid group should be
used instead of symmetric group. On the other hand,
it is known that inequivalent quantizations
of multiparticle systems on two-dimensional
manifolds are labeled by the irreducible
representations of the braid group,
giving rise to braid statistics \cite{leinaas,wilczek,wu}.
Being motivated by above-mentioned arguments,
in this paper we construct the Fock spaces of particles obeying
braid statistics.

Recent state of the problem is as follows. The Fock spaces
of the particles obeying statistics associated with
nontrivial representations of the symmetric group
have been constructed in \cite{minch}. The same has been
done in \cite{fiore} for the multiparticle systems
with the special emphasis on the quantum symmetrical
properties of systems under consideration. In the
number of papers Fock spaces are constructed, which
originate from the creation and annihilation operators
with nontrivial commutation relations being preserved
under the action of some quantum groups
\cite{kempf} (see also \cite{bied}-\cite{kul}).
The realization of the $q$-deformed Fock
space in terms of $q$-wedges have been done in
\cite{stern}.

In the present paper, a framework for constructing the
Fock space associated either
with any solution of the quantum Yang-Baxter equation
which obeys also the Hecke equation or with any representation
of the infinite dimensional Hecke algebra $H_{\infty}(q^2)$
is developed. Here we deal with the Hecke algebra, which is
a subalgebra of the group algebra of the braid group and,
on the other hand, is $q$-deformation of the usual
symmetric group. Other special realizations of the
braid group e.g., Birman-Wenzel-Murakami algebra or
truncated braid groups will be considered elsewhere.
Throughout this paper, $q$ is generic complex number with the exception
of root of unity.

Let us remind that the Hecke algebra $H_n(q^2)$ is defined
in terms of the generators $\sigma_1$, $\sigma_2$,
$\ldots$, $\sigma_{n-1}$,
which satisfy the following relations \cite{katriel,stern}:
\begin{eqnarray}
& &\sigma_i \sigma_{i+1} \sigma_i =
\sigma_{i+1} \sigma_{i} \sigma_{i+1}\;,\quad \quad
i=\overline{1,n-2}\;,\nonumber\\
& &\sigma_i\sigma_j=\sigma_j\sigma_i\;, \quad\mbox{if}
\quad |i-j|>1\;,\nonumber\\
& &\left(\sigma_i+1\right)\left(\sigma_i-q^{2} \right)=0, \quad
i=\overline{1,n-1}.
\label{ha}
\end{eqnarray}
First relation in (\ref{ha}) is (quantum)
Yang-Baxter equation and third one is
Hecke equation. The first and second relations
in (\ref{ha}) are defining
relations for the generators of the braid group. For
$q=1$ the relations (\ref{ha}) reduce to the generating relations
of the symmetric group while the Hecke algebra
reduces to the group algebra of the symmetric group.

There is a natural sequence of the Hecke algebras
$H_k(q^2)$, $k=\overline{2,n}$ :
\begin{equation}
H_2(q^2) \; \subset \; H_3(q^2) \; \subset \;
\ldots \; \subset H_n(q^2)\;,
\label{seq1}
\end{equation}
which are defined in terms of the generators
$\sigma_{n+1-k} \;,\; \ldots \;,\; \sigma_{n-1}$, respectively.
The fundamental
invariant of the Hecke algebra $H_k(q^2)$ is
\cite{katriel,dip}:
\begin{equation}
C_n^{(k)}=\sum\limits_{j=n+1-k}^{n-1}L_j \;, \quad k=\overline{2,n}\;,
\label{seq2}
\end{equation}
where $L_j$, $j=\overline{1,n-1}$, are the Murphy
operators \cite{dip}:
\begin{equation}
L_{n-1}=\sigma_{n-1}\;,
\label{seq3}
\end{equation}
\begin{equation}
L_j=\sum\limits_{k=j}^{n-1}q^{-2(k-j)}\sigma_k \sigma_{k-1} \ldots
\sigma_{j+1}\sigma_j\sigma_{j+1} \ldots \sigma_{k-1}\sigma_k \;,
\quad j=\overline{1,n-2}\;.
\label{seq4}
\end{equation}
Using (\ref{seq2})-(\ref{seq4}) and the Hecke equation, one
obtains \cite{katriel,dip} that $C_n^{(k)}$
belongs to the center of $H_k(q^2)$:
\begin{equation}
[C_n^{(k)},\sigma_j]=0 \;, \mbox{  for  } j=\overline{n+1-k,n-1} \;,
\label{seq5}
\end{equation}
and the relations hold
\begin{eqnarray}
& & [L_{j_1},L_{j_2}]=0\;, \quad j_1,j_2=\overline{1,n-1}
\nonumber \\
& & L_j=q^{-2}\sigma_j\, L_{j+1}\, \sigma_j + \sigma_j\;, \quad
j=\overline{1,n-2} \nonumber\\
& & L_j=C_n^{(n+1-j)}-C_n^{(n-j)}\;, \quad j=\overline{1,n-1}\;.
\label{seq6}
\end{eqnarray}

Using relations (\ref{ha})-(\ref{seq6}) one obtains that the
following propositions are valid.
\begin{pr}
The operators
\begin{equation}
P_S^{(n)}=\frac{1}{[n]_{q^2}!}\prod\limits_{j=1}^{n-1}
\left(1+L_j\right) \;,
\label{pr1}
\end{equation}
\begin{equation}
P_A^{(n)}=\frac{q^{n(n-1)}}{[n]_{q^2}!}\prod\limits_{j=1}^{n-1}
\left(1-q^{-2}L_j\right) \;,
\label{pr2}
\end{equation}
have the following properties:
\begin{equation}
\sigma_i P_S^{(n)}=q^2 P_S^{(n)} \;, \quad i=\overline{1,n-1} \;,
\label{pr3}
\end{equation}
\begin{equation}
\sigma_i P_A^{(n)}= -P_A^{(n)} \;, \quad i=\overline{1,n-1} \;,
\label{pr4}
\end{equation}
\begin{equation}
\left(P_S^{(n)}\right)^2=P_S^{(n)}\;,
\label{pr5}
\end{equation}
\begin{equation}
\left(P_A^{(n)}\right)^2=P_A^{(n)}\;,
\label{pr6}
\end{equation}
and
\begin{equation}
P_S^{(n)} \, P_A^{(n)}=P_A^{(n)} \, P_S^{(n)}=0\,.
\label{pr7}
\end{equation}
Here the following notations are used:
\begin{equation}
[n]_{q^2}=\frac{q^{2n}-1}{q^2-1}\;, \qquad [0]_{q^2}!=1,
\quad [n]_{q^2}!= \prod\limits_{k=1}^{n}[k]_{q^2}\,, \ n \geq 1.
\end{equation}
\end{pr}

\begin{pr}
Let $H_n(q^2)$ be algebra with involution $*$. Then
\begin{eqnarray}
\left(P_S^{(n)}\right)^{*}=P_S^{(n)}\;, \\
\mbox{and} \nonumber \\
\left(P_A^{(n)}\right)^{*}=P_A^{(n)}\;,
\label{pr8}
\end{eqnarray}
if one of the following conditions satisfies \\
1) $q^2$ is real number, $\overline{q^2}=q^2$ and
$\left(\sigma_i\right)^{*}=\sigma_i $, $i=\overline{1,n-1}$ \\
or \\
2) $\overline{q^2}=q^{-2}$ and
$\left(\sigma_i\right)^{*}=\left(\sigma_i\right)^{-1}$,
$i=\overline{1,n-1}$, where
\begin{equation}
\left(\sigma_i\right)^{-1}=q^{-2}\sigma_i +\left(q^{-2}-1\right)\;,
\label{pr9}
\end{equation}
according to Hecke equation.
\end{pr}
The above formulated propositions allow us to proceed to
construction of the braided Fock spaces.

Let ${\cal H}$ be either a separable Hilbert space or
even a finite dimensional
linear space with the orthonormal bases $\{e_k\}$,
$k \in {\cal I}$,
where ${\cal I}$ is a set of nonnegative integers,
${\cal I}= {\cal N} \bigcup \{0\}$, for the
first case and ${\cal I}=1,2, \ldots ,N$
($N$ is the dimension of $ \cal H$) for the
second one. Consider n-fold tensor power ${\cal H}^n={\cal H}^{\otimes n}$
with the orthonormal bases
$e_{k_1 k_2 \ldots k_n}=e_{k_1}\otimes e_{k_2}\otimes \ldots \otimes e_{k_n}$
and let ${\cal H}^0=C^1$. Then the direct sum
\begin{equation}
{\cal F}\left({\cal H}\right)=
\bigoplus\limits_{n=0}^{\infty}{\cal H}^n
\label{f1}
\end{equation}
is called the Fock space over ${\cal H}$.

Let us suppose that the action of the generators $\sigma_i$ of the Hecke
algebra $H_n(q^2)$ on the bases of ${\cal H}^n$, $n \geq 2$,
is given by the formula:
\begin{equation}
\sigma_i e_{k_1 k_2 \ldots k_n}=
\sum\limits_{l_i,l_{i+1} \in {\cal I}}
{\hat {\cal R}}\left(k_i, k_{i+1};l_i, l_{i+1}\right)
e_{k_1 k_2 \ldots k_{i-1} l_i l_{i+1} k_{i+2} \ldots k_n}\;.
\label{rm1}
\end{equation}
One can easy obtain from (\ref{ha}), (\ref{rm1}) that
${\hat {\cal R}}$-matrix satisfies the relations:
\begin{eqnarray}
& & \sum\limits_{l_1,l_2,l_3 \in {\cal I}}
{\hat {\cal R}}\left( m_1, m_2;l_1, l_2 \right)
{\hat {\cal R}}\left( l_2, m_3;l_3, n_3\right)
{\hat {\cal R}}\left( l_1, l_3;n_1, n_2\right)= \nonumber \\
& & =\sum\limits_{l_1,l_2,l_3 \in {\cal I}}
{\hat {\cal R}}\left( m_2, m_3;l_2, l_3\right)
{\hat {\cal R}}\left( m_1, l_2;n_1, l_1\right)
{\hat {\cal R}}\left( l_1, l_3;n_2, n_3\right)\;, \\
& & \sum\limits_{l_1,l_2 \in {\cal I}}
{\hat {\cal R}}\left(m_1, m_2;l_1, l_2\right)
{\hat {\cal R}}\left(l_1, l_2;n_1, n_2\right)=
(q^2-1){\hat {\cal R}}\left(m_1, m_2;n_1, n_2\right)+q^2\;,
\end{eqnarray}
(see, e.g. \cite{sudb} for the examples of such
${\hat {\cal R}}$-matrices).
For action of $\sigma_i$, on the
element $F^{(n)}$ of ${\cal H}^n$, $n \geq 2$
\begin{equation}
F^{(n)}=\sum\limits_{k_1, k_2, \ldots, k_n \in {\cal I}}
F^{(n)}\left(k_1, k_2, \ldots ,k_n\right)
e_{k_1 k_2 \ldots k_n}\;,
\label{fn1}
\end{equation}
one obtains from (\ref{rm1})
\begin{equation}
\sigma_i F^{(n)}= \sum\limits_{k_1, k_2, \ldots ,k_n \in {\cal I}}
e_{k_1 k_2 \ldots k_n}
\sum\limits_{l_i, l_{i+1} \in {\cal I}}
{\hat {\cal R}}\left(l_i, l_{i+1};k_i, k_{i+1}\right)
F^{(n)}\left(k_1,k_2,\ldots,k_{i-1},l_i,
l_{i+1},k_{i+2},\ldots,k_n\right)\;.
\label{rm2}
\end{equation}
The space of sequences $F$:
\begin{equation}
F=\left(F^{(0)}, F^{(1)}\left(k_1\right),
F^{(2)}\left(k_1, k_2\right), \ldots ,
F^{(n)}\left(k_1, k_2, \ldots, k_n\right), \ldots \right)\,
\label{fn2}
\end{equation}
with the scalar product
\begin{equation}
\left(F,{\tilde F} \right)=
\sum\limits_{n=0}^{\infty}
\left(F^{(n)},{\tilde F}^{(n)}\right)\;,
\label{fn3}
\end{equation}
\begin{equation}
\left(F^{(n)},{\tilde F}^{(n)}\right)=
\sum\limits_{k_1,k_2 \ldots ,k_n \in {\cal I}}
\overline{F^{(n)}\left(k_1, k_2, \ldots ,k_n\right)}
{\tilde F}^{(n)}\left(k_1, k_2, \ldots ,k_n\right)\;,
\label{fn4}
\end{equation}
is isomorphic to ${\cal F}\left({\cal H}\right)$ and will be
considered as ${\cal F}\left({\cal H}\right)$ henceforth.
The action of $\sigma_i$ on $F$ follows from (\ref{rm2}):
\begin{equation}
\left(\sigma_i F\right)^{(n)}\left( k_1, k_2, \ldots ,k_n\right)=
\sum\limits_{l_i, l_{i+1} \in {\cal I}}
{\hat {\cal R}}\left( l_i, l_{i+1};k_i, k_{i+1}\right)
F^{(n)}\left( k_1, k_2, \ldots, k_{i-1},
l_i, l_{i+1}, k_{i+2}, \ldots, k_n\right)\;,
\label{fn5}
\end{equation}
where $i<n$.

Let us turn now to the definition of the braided Fock spaces.
Setting
\begin{equation}
P_S= \bigoplus\limits_{n=0}^{\infty} P_S^{(n)}\;,
\label{proj1}
\end{equation}
and
\begin{equation}
P_A= \bigoplus\limits_{n=0}^{\infty} P_A^{(n)}\;,
\label{proj2}
\end{equation}
where
\begin{equation}
P_S^{(0)}=P_S^{(1)}=P_A^{(0)}=P_A^{(1)}=I\;,
\label{proj3}
\end{equation}
one obtains from
Proposition 1 that $P_S$ and
$P_A$ are projection operators (the operators
$P_S^{(n)}$, $P_S^{(n)}$ for $n \geq 2$ are defined by
formulas (\ref{pr1}), (\ref{pr2}) of Proposition 1).
Furthermore, if one defines the action of involution $*$
by adjontion, i.e.
\begin{equation}
\left(\sigma_i \right)^{*}=\left(\sigma_i \right)^{\dagger}\;,
\label{proj4}
\end{equation}
where $\left(\sigma_i \right)^{\dagger}$ is adjoint
of $\sigma_i$, then it follows from Proposition 2
that $P_S$ and $P_A$ are orthogonal projectors, if
either $q^2$ is real number and $\sigma_i$ is selfadjoint
operator or $\overline{q^2}=q^{-2}$ and $\sigma_i$ is unitary
operator. In what follows it will be assumed that the conditions
of Proposition 2 hold.

The braided Fock spaces we are looking for are
now defined by
\begin{equation}
{\cal F}_S\left({\cal H}\right)=
P_S{\cal F}\left({\cal H}\right)\;,
\label{bfss1}
\end{equation}
\begin{equation}
{\cal F}_A\left({\cal H}\right)=
P_A{\cal F}\left({\cal H}\right)\;.
\label{bfsa1}
\end{equation}
According to Proposition 1, the
${\cal F}_S\left({\cal H}\right)$
and ${\cal F}_A\left({\cal H}\right)$ are the spaces of sequences
\begin{equation}
F_S=\left(F_S^{(0)}, F^{(1)}_S\left(k_1\right),
F^{(2)}_S\left(k_1, k_2\right), \ldots ,
F^{(n)}_S\left(k_1, k_2, \ldots, k_n\right), \ldots \right)\;,
\label{bfss2}
\end{equation}
\begin{equation}
F_A=\left(F_A^{(0)}, F^{(1)}_A\left(k_1\right),
F^{(2)}_A\left(k_1, k_2\right), \ldots ,
F^{(n)}_A\left(k_1, k_2, \ldots, k_n\right), \ldots \right)\;,
\label{bfsa2}
\end{equation}
respectively, where
$F^{(n)}_S\left(k_1, k_2, \ldots, k_n\right)$
for $n \geq 2$ is an
eigenelement of $\sigma_i$, $i=\overline{1,n-1}$
with eigenvalue $q^2$ and
$F^{(n)}_A\left(k_1, k_2, \ldots, k_n\right)$
for $n \geq 2$ is an
eigenelement of $\sigma_i$, $i=\overline{1,n-1}$
with eigenvalue $-1$. The above-mentioned means
that ${\cal F}_S\left({\cal H}\right)$
and ${\cal F}_A\left({\cal H}\right)$ are Fock spaces
of particles obeying braid statistics associated with
certain solution ${\hat {\cal R}}$ of the Yang-Baxter equation.

To construct annihilation and creation operators on
${\cal F}_S\left({\cal H}\right)$ and
${\cal F}_S\left({\cal H}\right)$, let us
first introduce on
${\cal F}\left({\cal H}\right)$ the
operator $b(k)$, $k \in {\cal I}$ \cite{reed} by its
action on $F \in {\cal F}\left({\cal H}\right)$:
\begin{eqnarray}
& & \left( b(k)\; F\right)^{(0)}=0 \;, \nonumber \\
& & \left( b(k)\; F\right)^{(n)}
\left( k_1,k_2, \ldots ,k_n \right)=
F^{(n+1)}
\left( k,k_1,k_2, \ldots ,k_n \right)\;, \quad n \geq 1\;.
\label{a1}
\end{eqnarray}
The adjoint of $b(k)$, operator $b^{\dagger}(k)$ acts on
${\cal F}\left({\cal H}\right)$ by
\begin{equation}
\left( b^{\dagger}(k)\;F\right)^{(n)}
\left(k_1,k_2, \ldots ,k_n \right)=
\delta_{k k_1} F^{(n-1)}
\left(k_2, \ldots ,k_n\right)\;.
\label{a2}
\end{equation}
Let us introduce following operators
for the spaces ${\cal F}_S\left({\cal H}\right)$
and ${\cal F}_A\left({\cal H}\right)$ \cite{reed}:
\begin{equation}
b_S(k)=P_S\;b(k)\;P_S\;, \quad
b_S^{\dagger}(k)=P_S\; b^{\dagger}(k)\;P_S\;,
\label{a3}
\end{equation}
\begin{equation}
b_A(k)=P_A\;b(k)\;P_A\;, \quad
b_A^{\dagger}(k)=P_A\;b^{\dagger}(k)\;P_A\;.
\label{a4}
\end{equation}
The action of the operators
$b_S(k)$ and $b_A(k)$ is given by
\begin{eqnarray}
& & \left(b_S(k)\;F_S\right)^{(0)}=0\;, \nonumber \\
& & \left(b_S(k)\;F_S\right)^{(n)}
\left(k_1,k_2, \ldots,k_n\right)=
F_S^{(n+1)}\left(k,k_1,k_2, \ldots, k_n\right)\;, \quad
n \geq 1\;,
\label{a5}
\end{eqnarray}
\begin{eqnarray}
& & \left(b_A(k)\;F_A\right)^{(0)}=0\;, \nonumber \\
& & \left(b_A(k)\;F_A\right)^{(n)}
\left(k_1,k_2, \ldots,k_n\right)=
F_A^{(n+1)}\left(k,k_1,k_2, \ldots, k_n\right)\;, \quad
n \geq 1\;.
\label{a6}
\end{eqnarray}
For the operators $b^{\dagger}_S(k)$ and $b^{\dagger}_A(k)$
formulas are more complicated:
\begin{eqnarray}
& & \left(b^{\dagger}_S(k)\;F_S\right)^{(n)}
\left(k_1,k_2, \ldots,k_n\right)=
\left(\ \frac{1}{\left[n\right]_{q^2}}\ \right)
\bigg\{ \delta_{kk_1}F_S^{(n-1)}
\left(k_2, \ldots, k_n\right)+ \nonumber \\
& & +\sum\limits_{l_2 \in {\cal I}}
{\hat {\cal R}}\left( k, l_2;k_1, k_2 \right)
F_S^{(n-1)}\left(l_2, k_3,\ldots, k_n\right)+ \nonumber \\
& & +\sum\limits_{l_2, l_3, j_2 \in {\cal I}}
{\hat {\cal R}}\left(l_2, l_3;k_2, k_3 \right)
{\hat {\cal R}}\left( k, j_2;k_1, l_2 \right)
F_S^{(n-1)}\left(j_2, l_3, k_4, \ldots, k_n\right)+ \nonumber \\
& & +\sum\limits_{m=3}^{n-1}
\sum\limits_{l_2, \ldots, l_{m+1}, j_2, \ldots, j_m \in {\cal I}}
{\hat {\cal R}}\left(l_m, l_{m+1};k_m, k_{m+1} \right)
\left(\prod\limits_{s=3}^{m}
{\hat {\cal R}}\left(l_{s-1}, j_s;k_{s-1}, l_s \right)\right)
\times \nonumber \\
& & \times {\hat {\cal R}}\left(k, j_2;k_1, l_2 \right)
F_S^{(n-1)}\left(j_2, \ldots, j_m, l_{m+1},
k_{m+2}, \ldots, k_n\right)\bigg\}\;,
\label{a7}
\end{eqnarray}
\begin{eqnarray}
& & \left(b^{\dagger}_A(k)\;F_A\right)^{(n)}
\left(k_1,k_2, \ldots,k_n\right)=
\left(\ \frac{q^{2(n-1)}}{\left[n\right]_{q^2}}\ \right)
\bigg\{ \delta_{kk_1}F_A^{(n-1)}
\left(k_2, \ldots, k_n\right)- \nonumber \\
& & -q^{-2}\sum\limits_{l_2 \in {\cal I}}
{\hat {\cal R}}\left( k, l_2;k_1, k_2 \right)
F_A^{(n-1)}\left(l_2, k_3,\ldots, k_n\right)+ \nonumber \\
& & +q^{-4}\sum\limits_{l_2, l_3, j_2 \in {\cal I}}
{\hat {\cal R}}\left(l_2, l_3;k_2, k_3 \right)
{\hat {\cal R}}\left( k, j_2;k_1, l_2 \right)
F_A^{(n-1)}\left(j_2, l_3, k_4, \ldots, k_n\right)+ \nonumber \\
& & +\sum\limits_{m=3}^{n-1}\left(-q^{-2}\right)^m
\sum\limits_{l_2, \ldots, l_{m+1}, j_2, \ldots, j_m \in {\cal I}}
{\hat {\cal R}}\left(l_m, l_{m+1};k_m, k_{m+1} \right)
\left(\prod\limits_{s=3}^{m}
{\hat {\cal R}}\left(l_{s-1}, j_s;k_{s-1}, l_s \right)\right)
\times \nonumber \\
& & \times {\hat {\cal R}}\left(k, j_2;k_1, l_2 \right)
F_A^{(n-1)}\left(j_2, \ldots, j_m, l_{m+1},
k_{m+2}, \ldots, k_n\right)\bigg\}\;.
\label{a8}
\end{eqnarray}
 From equations (\ref{a5})-(\ref{a8}) one gets:
\begin{eqnarray}
& & \left(\left\{b_S(k)\,b_S^{\dagger}(k')-
\frac{\left[ N \right]_{q^2}}{\left[ N+1 \right]_{q^2}}
\sum\limits_{l,l' \in {\cal I}}
{\hat {\cal R}}\left(k', l;k, l' \right)
b_S^{\dagger}(l')\,b_S(l)\right\}F_S\right)^{(n)}
\left(k_1,k_2, \ldots, k_n\right)= \nonumber \\
& & =\frac{1}{\left[ n+1 \right]_{q^2}}
\delta_{k\;k'}\,F_S^{(n)}
\left(k_1,k_2, \ldots, k_n\right)\;,
\label{a9}
\end{eqnarray}
\begin{eqnarray}
& & \left(\left\{b_A(k)\,b_A^{\dagger}(k')+
\frac{\left[ N \right]_{q^2}}{\left[ N+1 \right]_{q^2}}
\sum\limits_{l,l' \in {\cal I}}
{\hat {\cal R}}\left(k', l;k, l' \right)
b_A^{\dagger}(l')\,b_A(l)\right\}F_A\right)^{(n)}
\left(k_1,k_2, \ldots, k_n\right)= \nonumber \\
& & =\frac{q^{2n}}{\left[ n+1 \right]_{q^2}}
\delta_{k\;k'}\,F_A^{(n)}
\left(k_1,k_2, \ldots, k_n\right)\;,
\label{a10}
\end{eqnarray}
\begin{eqnarray}
& & \left(\left\{b_S^{\dagger}(k)\,b_S^{\dagger}(k')-
q^{-2}\sum\limits_{l,l' \in {\cal I}}
{\hat {\cal R}}\left(k, k';l, l' \right)
b_S^{\dagger}(l)\,b_S^{\dagger}(l')\right\}F_S\right)^{(n)}
\left(k_1,k_2, \ldots, k_n\right)= 0\;, \nonumber \\
& & \left(\left\{b_S(k)\,b_S(k')-
q^{-2}\sum\limits_{l,l' \in {\cal I}}
{\hat {\cal R}}\left(l', l;k', k \right)
b_S(l)\,b_S(l')\right\}F_S\right)^{(n)}
\left(k_1,k_2, \ldots, k_n\right)= 0\;,
\label{a11}
\end{eqnarray}
\begin{eqnarray}
& & \left(\left\{b_A^{\dagger}(k)\,b_A^{\dagger}(k')+
\sum\limits_{l,l' \in {\cal I}}
{\hat {\cal R}}\left(k, k';l, l' \right)
b_A^{\dagger}(l)\,b_A^{\dagger}(l')\right\}F_A\right)^{(n)}
\left(k_1,k_2, \ldots, k_n\right)= 0\;, \nonumber \\
& & \left(\left\{b_A(k)\,b_A(k')+
\sum\limits_{l,l' \in {\cal I}}
{\hat {\cal R}}\left(l', l;k', k \right)
b_A(l)\,b_A(l')\right\}F_A\right)^{(n)}
\left(k_1,k_2, \ldots, k_n\right)= 0\;.
\label{a12}
\end{eqnarray}
In (\ref{a9}), (\ref{a10}) $N$ is the operator of a number of
particles, which acts on ${\cal F}\left({\cal H}\right)$ by
\begin{equation}
\left(N\; F\right)^{(n)}
\left(k_1,k_2, \ldots, k_n\right)=
n\;F^{(n)}
\left(k_1,k_2, \ldots, k_n\right)\;.
\end{equation}

It is interesting to note that equations (\ref{a9})-(\ref{a12})
are obtained from relations (\ref{a5})-(\ref{a8})
without any assumptions about selfadjointness of
the projectors $P_S$ and $P_A$.

Now one can define the creation and annihilation operators
for the spaces ${\cal F}_S\left({\cal H}\right)$ and
${\cal F}_A\left({\cal H}\right)$:
\begin{eqnarray}
& & a_S(k)=\sqrt{\left[ N+1 \right]_{q^2}}b_S(k)\;,
\nonumber \\
& & a_S^{\dagger}(k)=
b_S^{\dagger}(k)\sqrt{\left[ N+1 \right]_{q^2}}\;,
\label{a13}
\end{eqnarray}
\begin{eqnarray}
& & a_A(k)=q^{-N}\sqrt{\left[ N+1 \right]_{q^2}}b_A(k)\;,
\nonumber \\
& & a_A^{\dagger}(k)=
b_A^{\dagger}(k)q^{-N}\sqrt{\left[ N+1 \right]_{q^2}}\;.
\label{a14}
\end{eqnarray}
The operators $a_S(k)$ and $a_S^{\dagger}(k)$
are adjoint if the second condition of the Proposition 2
satisfies. The same is valid for the operators
$a_A(k)$ and $a_A^{\dagger}(k)$.
Following commutation relations do not depend on the
above-mentioned condition of adjointness:
\begin{equation}
a_S(k)\,a_S^{\dagger}(k')-
\sum\limits_{l,l' \in {\cal I}}
{\hat {\cal R}}\left(k', l;k, l' \right)
a_S^{\dagger}(l')\,a_S(l)=
\delta_{k\;k'}\;,
\label{a15}
\end{equation}
\begin{equation}
a_S^{\dagger}(k)\,a_S^{\dagger}(k')-
q^{-2}\sum\limits_{l,l' \in {\cal I}}
{\hat {\cal R}}\left(k, k';l, l' \right)
a_S^{\dagger}(l)\,a_S^{\dagger}(l')= 0\;,
\label{a16}
\end{equation}
\begin{equation}
a_S(k)\,a_S(k')-
q^{-2}\sum\limits_{l,l' \in {\cal I}}
{\hat {\cal R}}\left(l', l;k', k \right)
a_S(l)\,a_S(l')=0\;,
\label{a17}
\end{equation}
\begin{equation}
a_A(k)\,a_A^{\dagger}(k')+ q^{-2}
\sum\limits_{l,l' \in {\cal I}}
{\hat {\cal R}}\left(k', l;k, l' \right)
a_A^{\dagger}(l')\,a_A(l)=\delta_{k\;k'}\;,
\label{a18}
\end{equation}
\begin{equation}
a_A^{\dagger}(k)\,a_A^{\dagger}(k')+
\sum\limits_{l,l' \in {\cal I}}
{\hat {\cal R}}\left(k, k';l, l' \right)
a_A^{\dagger}(l)\,a_A^{\dagger}(l')= 0\;,
\label{a19}
\end{equation}
\begin{equation}
a_A(k)\,a_A(k')+
\sum\limits_{l,l' \in {\cal I}}
{\hat {\cal R}}\left(l', l;k', k \right)
a_A(l)\,a_A(l')= 0\;.
\label{a20}
\end{equation}
Operators $a_S(k)$ ($a_A(k)$) and $a_S^{\dagger}(k)$
($a_A^{\dagger}(k)$) with commutation relations (\ref{a15})-
(\ref{a17}) ((\ref{a18})-(\ref{a20})) are known as quantum
deformed oscillator algebra.

Relations (\ref{a16}), (\ref{a19}) are preserved under the
linear transformations
\begin{equation}
a^{\dagger}(k) \to
\sum\limits_{l \in {\cal I}}T(k,l)\,a^{\dagger}(l),
\quad k \in {\cal I}\;,
\label{a21}
\end{equation}
where $a^{\dagger}(k)$ denote either $a_S^{\dagger}(k)$
or $a_A^{\dagger}(k)$ and $T(i,j)$, $i,j \in {\cal I}$ obey
relations:
\begin{equation}
\sum\limits_{p,s \in {\cal I}}T(i,p)T(j,s)
{\hat {\cal R}}\left(p, s;k, l \right)=
\sum\limits_{p,s \in {\cal I}}
{\hat {\cal R}}\left(i, j;p, s \right)
T(p,k)T(s,l)\;,
\label{a22}
\end{equation}
\begin{equation}
T(i,j)a^{\dagger}(k)=a^{\dagger}(k)T(i,j)\;, \quad
i,j,k \in {\cal I}\;.
\label{a23}
\end{equation}
Similarly, relations (\ref{a17}), (\ref{a20}) are preserved
under the linear transformations
\begin{equation}
a(k) \to
\sum\limits_{l \in {\cal I}}T^{*}(k,l)\,a(l),
\quad k \in {\cal I}\;,
\label{a24}
\end{equation}
where $a(k)$ denote either $a_S(k)$
or $a_A(k)$ and $T^{*}(i,j)$, $i,j \in {\cal I}$ obey
relations:
\begin{equation}
\sum\limits_{p,s \in {\cal I}}T^{*}(i,p)T^{*}(j,s)
{\hat {\cal R}}\left(l, k;s, p \right)=
\sum\limits_{p,s \in {\cal I}}
{\hat {\cal R}}\left(s, p;j, i \right)
T^{*}(p,k)T^{*}(s,l)\;,
\label{a25}
\end{equation}
\begin{equation}
T^{*}(i,j)a(k)=a(k)T^{*}(i,j)\;, \quad
i,j,k \in {\cal I}\;.
\label{a26}
\end{equation}
One has not mixed designation $*$ in (\ref{a25}), (\ref{a26}) with
the involution of the Hecke algebra introduced in Proposition 2.

Relations (\ref{a22}) define quantum algebra associated with
${\hat {\cal R}}$-matrix
${\hat {\cal R}}\left(i, j;k, l \right)$ while relations
(\ref{a25}) define quantum algebra associated with
transposed ${\hat {\cal R}}$-matrix
${\hat {\cal R}}^{*}\left(i, j;k, l \right)=
{\hat {\cal R}}\left(l, k;j, i \right)$.

Let us close this paper with some concluding remarks.
Starting from the representation (\ref{rm1}), (\ref{fn5})
of the Hecke
algebra, we built the Fock space of particles, which obey
the braid statistics associated with a certain solution
${\hat {\cal R}}$ of the Yang-Baxter equation. Braid
statistics nature of the Fock spaces under consideration is
obvious from the commutation relations (\ref{a15})-(\ref{a20}).
It follows from Propositions 1 and 2 that the
framework developed in the paper allows us to build
Fock spaces associated with infinite dimensional Hecke
algebra $H_{\infty}(q^2)$ acting on Fock space over ${\cal H}$,
${\cal F}\left({\cal H}\right)$ instead of one associated with
${\hat {\cal R}}$.

It is interesting to note that there does not exist nontrivial
deformation of the algebra of oscillators within the category
of algebras \cite{pillin} at least for a finite dimensional case.
It means that any deformation of the
oscillator algebra can be considered as a deformation
map from a undeformed algebra. There exist number of examples
for such deformation maps \cite{milek}. From the other hand,
being separable Hilbert spaces,
the Fock spaces constructed in this paper and
usual (anti)symmetric Fock space are isomorphic. It seems
to be interesting to investigate connections between such
isomorphisms and the above-mentioned deformation map.
Another interesting problem arises due to a so called
quon algebra \cite{green}, for which the associated Fock space
can be constructed also. To obtain such a Fock space
as one associated with the braid group is a task of future
investigations.

The author thanks A.I.Bugrij and V.N.Shadura for useful
discussions.

\end{document}